\documentclass[twocolumn,amsmath,amssymb]{revtex4-1}%
\usepackage{natbib}
\usepackage{graphicx}
\usepackage{hyperref}
\usepackage{comment}
\usepackage{float}

\begin{document}
\title{The Second International Pulsar Timing Array Mock Data Challenge}
%\author{Jeffrey S.\ Hazboun} 
\author{Jeffrey S Hazboun}\affiliation{Physical Sciences Division, University of Washington Bothell, 18115 Campus Way NE, Bothell, WA 98011}\email{hazboun@uw.edu}
\author{Chiara M. F. Mingarelli}\affiliation{Center for Computational Astrophysics, Flatiron Institute, 162 Fifth Ave, New York, NY 10010}\email{cmingarelli@flatironinstitute.org}
\author{Kejia Lee}\affiliation{Kavli Institute for Astronomy and Astrophysics, Peking University, Beijing 100871, P.R.China}\email{kjlee@pku.edu.cn}
%\author{on behalf of the\\ IPTA Data Challenge Working Group}

\begin{abstract}
\noindent
The International Pulsar Timing Array (IPTA) is a galactic-scale gravitational-wave observatory that monitors an array of millisecond pulsars. The timing precision of these pulsars is such that one can measure the correlated changes in pulse arrival times accurately enough to search for the signature of a stochastic gravitational-wave background. As we add more pulsars to the array, and extend the length of our dataset, we are able to observe at ever lower gravitational-wave frequencies. As our dataset matures we are approaching a regime where a detection is expected, and therefore testing current data analysis tools is crucial, as is the development of new tools and techniques.
In this spirit, here we introduce the second IPTA Mock Data Challenge, and briefly review the first. The purpose of this challenge is to foster the development of detection tools for pulsar timing arrays and to cultivate interaction with the international gravitational-wave community. IPTA mock datasets can be found at the IPTA GitHub page, \url{https://github.com/ipta/mdc2}.
\end{abstract}

\maketitle

\section{The International Pulsar Timing Array}
The International Pulsar Timing Array (IPTA) \cite{Hobbs:2009yy, manchester:2013, mclaughlin:2014, VerbiestEtAl:2016, lentatietal:2016, caballeroetal:2018} is a consortium of constituent pulsar timing array (PTA) collaborations, working together across continents. It includes a large and diverse group of scientists spanning many disciplines, from hardware development to tests of General Relativity,  dedicated to the search for nanohertz gravitational waves (GWs). The founding members include the Parkes Pulsar Timing Array~\cite{Shannon:2015ect, reardonetal:2016}, the European Pulsar Timing Array~\cite{Lentati:2015qwp, desvignes+:2016, bps+15, TaylorEtAl:2015, caballeroetal:2016}, and the North American Nanohertz Observatory for Gravitational Waves (NANOGrav), see e.g. \cite{ArzoumanianEtAl:2018, Arzoumanian:2018saf, gentileetal:2018}. 

Currently, the IPTA is finalizing preparations on its second data release, and is looking to expand to include new and emerging PTA collaborations, like the Indian Pulsar Timing Array and the Chinese Pulsar Timing Array, in a joint effort to observe low-frequency GWs.

\section{Pulsar Timing Arrays as Gravitational Wave Detectors}
Pulsar timing uses the precision timing of ultra-stable millisecond pulsars to search for nanohertz GWs.  
%The arrival time of pulses from a pulsar are monitored on Earth, and compared against a model for the expected arrival times. 
GWs sweeping through the galaxy advance or delay the arrival times of these regular pulses from the pulsars. An advance or delay in a single pulsar could be due to intrinsic noise processes in the pulsar, or other unmodeled effects on the pulse as it travels through the galaxy, however a GW signal would be present in all the pulsar residuals -- the difference between the expected and the actual arrival time. In the presence of an isotropic GW background (GWB), likely formed from the cosmic merger history of supermassive black holes (SMBHs), these residuals will be correlated, with the correlation function described by the Hellings and Downs curve \cite{Hellings:1983fr}.

The primary source of low-frequency GWs is expected to be from early the inspiral phase of supermassive black hole binaries (SMBHBs), with masses greater than $\sim 10^8~M_\odot$. The incoherent superposition of these GWs should form a GWB with an amplitude $A$ ranging anywhere from $1\times 10^{-16}-2\times 10^{-15}$, at a reference frequency of $1/$yr~\cite{RR95, wl03,jb03, s13, mop14, Shankaretal:2016, ryuetal:2018, bonnettietal:2018}. The amplitude of the GWB depends strongly on the astrophysics underlying SMBH mergers: the merger rate (or stalling), the black hole masses, and the SMBH population fraction (how many galaxies host SMBHs). The shape of the spectrum can also be affected by stars and gas interacting with the binary, as well as binary eccentricity, all of which cause the strain spectrum to turn over at low frequencies \cite{s13b, scm15, ArzoumanianEtAl:2016, Arzoumanian:2018saf}. As such, the most up-to-date analyses of PTA data have already set astrophysically interesting bounds on these processes \cite{ Shannon:2015ect, ArzoumanianEtAl:2016, Arzoumanian:2018saf}. 

Since PTAs are sensitive to any class of nHz GW signal, they can also provide new insights into cosmology -- e.g. limits on the tensor-to-scalar ratio $r$ and the tensor index $n_t$~\cite{lms+15,Lentati:2015qwp, Arzoumanian:2018saf}, tests of general relativity \cite{ss12, cot+18}, and limits on cosmic string tension ~\cite{sbs12, Lentati:2015qwp, ArzoumanianEtAl:2016, Arzoumanian:2018saf, bos18}. 

The strongest signal for PTAs is expected to be an isotropic GWB, as described above~\cite{rsg15}. It is expected to be detected in the next five years, though the amplitude of the GWB depends on the astrophysics underlying the cosmic merger history of SMBHs~\cite{sejr13, tve+16, vs16}. Single-source GWs from local SMBHBs are expected to be detected in the next 10 years, with similar caveats \cite{mls+17}. 

\begin{figure*}[ht]
\begin{center}
\includegraphics[scale=.6]{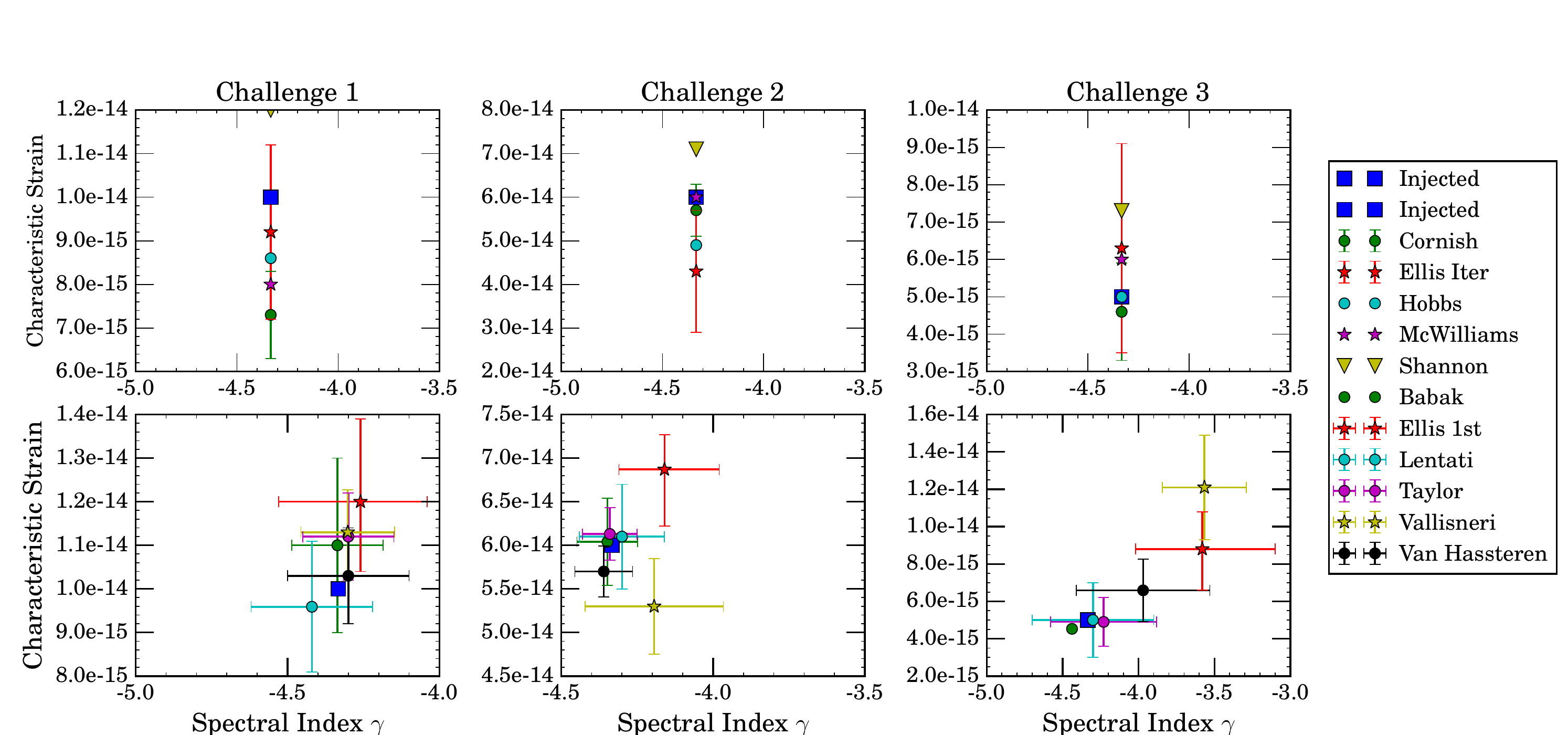}		
\caption{Submitted results from the first IPTA MDC, \citeauthor{esc12, vHmv+13, tgl13, c12}. Shown here are the limits on $A_c$, the amplitude of the GWB at a reference frequency of $1/$yr, and the index of the strain spectrum, $\gamma$. The upper panels are for searches with fixed $\gamma$, and the lower panels search for constraints on $A_c$ and $\gamma$. For a GWB created by SMBHBs, $\gamma = -4.33$. Error bars are $1\textnormal{-}\sigma$.}
\label{fig:mdc_submissions}
\end{center}
\end{figure*}

\section{First Mock Data Challenge}
In order to encourage development of GW search algorithms for pulsar timing data, the IPTA issued the first mock data challenge (MDC) in 2012. This first MDC was developed and administered by Fredrick Jenet, Kejia Lee and Michael Keith.  The challenge solicited 13 analysis code submissions and spawned a number of publications from those submissions, e.g. \cite{esc12, vHmv+13, tgl13, tgl12, c12}. The original challenge contained two rounds, each with 3 datasets. One of the rounds was open, with all of the noise characteristics and signal injections known to the participants. The other dataset was closed, and was the focus of the submissions. The injections into all datasets were relatively strong signals excluded by recent upper limits set in the literature by various PTAs. A summary of the submissions for the first MDC can be seen in Figure~\ref{fig:mdc_submissions}. The original MDC datasets, and results, can be found at: {\footnotesize \url{ http://www.ipta4gw.org/?page_id=89}}

\section{Sources of Noise in PTAs}

Since their discovery, pulsars have been used as probes of a number of interesting astrophysical and gravitational phenomena, including orbital decay due to gravitational waves \cite{Taylor:1982zz}, probes of the bulk movement and location of the interstellar medium \cite{Cordes:2002wz}, finding exoplanets \cite{Wolszczan:1992zg} and tests of the equivalence principle \cite{Archibald:2018oxs}. 
These, and the many other astrophysical effects that change the propagation of radio pulses from pulsars, are noisy additions from the perspective of gravitational wave astronomy.  

Noise characterization and mitigation is an active area of research in the pulsar timing community. A full review of the subject is out of the scope of this work, but participants in the IPTA MDC are encouraged to take a look into the extensive literature on the subject. The following manuscripts, and references therein, are a nice starting point, \cite{Lam:2017ysu, Cordes:2010fh, Lam:2015vif, Lam:2016iie, Shannon:2016fez, vs18}.

\section{IPTA Mock Data Challenge 2}
Over the next few months the IPTA will release two mock data challenges. The first (MDC2) is being released now, while the next release (MDC3) will happen in a few months near the end of 2018. 

MDC2 contains two separate groups of datasets. These datasets are designed as an introduction to analyzing pulsar timing data and are similar in complexity to the original MDC1. MDC3 will contain more realistic signal and noise injections.

Within MDC2, Group 1 each dataset contains a version of the data which is evenly spaced (a-version), and another set which has a more realistic, unevenly spaced cadence of observation epochs (b-version). Both the signal and noise injections for these datasets are identical, and the evenly-spaced dataset is provided as a simple first problem for students as well as a comparison of how unevenly-sampled data complicates PTA data analysis. Group 1 is made up of open datasets. Group 1, Dataset 1 contains only white noise, i.e. TOA measurement uncertainties, for the individual pulsars and only one signal; a stochastic gravitational wave background injected as appropriate spatial-correlations between the various pulsars.  Group 1, Dataset 2 also contains a stochastic GWB and white noise, but additionally has red noise injections for the pulsars. Red noise here refers to a low-frequency process that produces errors in data that are correlated over long timescales. This red noise is modeled in the Fourier domain where the power spectral density is modeled as a power law with specific spectral index. Details about the power-law amplitude and spectral index for each pulsar can be found in the specific dataset notes. Group 1, Dataset 3 contains no stochastic background, but does contain a single SMBHB source. 

MDC2 Group 2 has a similar signal/noise content as Group 1, however there is no longer an evenly sampled version of the data. 
These datasets are closed in the sense that we do not reveal which dataset includes which noise and signal types. 
However, we do reveal that the dataset includes three combinations of signal. 
One solely contains a stochastic GWB, one contains a stochastic GWB and a single SMBHB source and one contains no stochastic background but two SMBHB signals. 
All contain both white noise and red noise. 

Two groups of three datasets will be released near the end of 2018 in MDC3.

\subsection{Format of Datasets}
In Table~\ref{tab:dataset_summary} we summarize the contents of the 6 different datasets included in MDC2. Details about the noise parameters injected into Group~1 can be found along with the datasets on the IPTA GitHub page in the form of a {\tt json} file. 

The datasets included in the MDC include parameter files containing the timing model parameters for a pulsar, and timing files containing the integrated times-of-arrival or TOAs of pulsar pulses. 
These files are referred to as {\tt par} and {\tt tim} files, respectively.
The {\tt par} files contain the relevant parameters needed to construct such a model to remove these effects, but its construction is left to the participant. 

\begin{table}[h]
    \centering
    \begin{tabular}{c|c|c|c|c|c}
    Group. \\Dataset  & Time Span & Freq (GHz) & Cadence & Noise & Signals\\
    \hline
     g1.d1a(b) & 15 yrs & 1.44  & 30 days & WN & SB\\
     g1.d2a(b) & 15 yrs & 1.44  & 30 days & WN,RN & SB \\
     g1.d3a(b) & 15 yrs & 1.44  & 30 days & WN & SB+SS\\
     g2.d1 & 15 yrs & 0.8,1.44  & 30 days & - & -\\
     g2.d2 & 15 yrs & 0.8,1.44  & 30 days & - & - \\
     g2.d3 & 15 yrs & 0.8,1.44  & 30 days & - & - \\
    \end{tabular}
    \caption{Dataset formats are summarized above for the 6 datasets included in MDC2. All datasets in Group 1 have an a) and b) version corresponding to the evenly sampled and unevenly sampled versions. Here WN \& RN are short for white noise and red noise, respectively. The abbreviations in the last column are Stochastic Background (SB) and Single Source (SS). The last two columns for Group 2 are left empty since the dataset is closed.}
    \label{tab:dataset_summary}
\end{table}

\section{Pulsar Software}
There is a rich history in the pulsar timing community of scientist-written and maintained code bases useful for timing pulsars and, more recently, for doing gravitational wave analyses. 
The TOAs in the {\tt tim} file contain Doppler shifts from the motion of the Earth in its orbit, relative proper motion between the solar system and the pulsar system and any orbital motion of the pulsar, if in a binary system. 
In addition, various clock corrections, observatory location coordinates and other observation dependent effects must be taken into account to build an adequate timing model for a given pulsar. 
While the participant is free to write their own code, there are three publicly available sets of software for modeling pulsar TOAs. 
The first is {\tt TEMPO}\cite{Nice:2015a}, written and distributed in {\tt fortran}, with code found at \footnote{ \url{http://tempo.sourceforge.net/} }. 
A different code, originally based on the same algorithms as {\tt TEMPO} is {\tt TEMPO2} \cite{Hobbs:2006_mnras}, written and distributed in {\tt c++}, with code found at \footnote{ \url{https://bitbucket.org/psrsoft/tempo2} }. 
Even more recent is {\tt PINT}, \footnote{ \url{https://github.com/nanograv/pint}}, written and distributed in Python. 
There is also a publicly available Python wrapper for {\tt TEMPO2} called {\tt libstempo} \footnote{\url{https://github.com/vallis/libstempo}}. 
While new software development is welcome, the participant is encouraged to use one of these existing packages for building a pulsar timing model, and to focus their time on the data analysis algorithm development. 

The existence of a number of well developed PTA gravitational wave data analysis packages needs to be recognized and commented on here. 
Among the many reasons for holding a MDC, two are relevant when discussing these previous code bases; to encourage new researchers to become involved with PTA data analysis and to foster the cross pollination of analysis techniques. 
In order to facilitate the first of these, the development of an entirely independent code for PTA analysis is not required for participation in the MDC, though obviously welcome and encouraged. 

Besides the existing analysis codes ({\tt TempoNest} \cite{Lentati:2013rla}, {\tt 42}, {\tt PAL2} \cite{Ellis:2017a}, {\tt NX01} \footnote{\url{https://zenodo.org/record/250258#.W4hT05NKhBw}}, {\tt piccard} \footnote{\url{https://github.com/vhaasteren/piccard}}) there also exists a code built as a framework for constructing a PTA analysis developed by the NANOGrav collaboration \footnote{While many of the techniques and sub-models within {\tt enterprise} are based on earlier analysis codes, the structure of {\tt enterprise} is designed to allow for the user to plugin various sub-models of their own into the analysis.}, called {\tt enterprise} \footnote{\url{https://github.com/nanograv/enterprise}}. 
Since these tools are accessible widely, we do not prohibit their use in the MDC, but will categorize them differently when discussing them in any manuscript that summarizes the MDC. 
Entries based on an existing code base (unless submitted by the developer of that code) must add some significant new ability to the base code in order to warrant their submission, and, of course, cite the original developer. An example of such an addition could be a new way of modeling noise in a pulsar, or using the output from the original code to calculate a different statistic.

Please see \url{https://github.com/ipta/mdc2} for the datasets and ongoing updates about the MDC. 
Direct any questions to \url{hazboun@uw.edu}. 

\bibliography{mdc}
\bibliographystyle{apsrev4-1}

\end{document}